%% file: main.tex
\documentclass[a4paper, 11pt]{article}
\usepackage{pos}
\usepackage[utf8]{inputenc}
\usepackage{amsmath}
\usepackage{graphicx}
\usepackage[english]{babel}
\usepackage{hyperref}
\usepackage{doi}
\usepackage{multicol}
\newcommand{\etal}{\textit{et al.~}}

\newcommand{\EGeV}{E_{4\mathrm{GeV}}}

\title{Implications of turbulence-dependent diffusion on cosmic-ray spectra}

\author*[a,b]{J. Dörner}
\author[a,b,c]{P. Reichherzer}
\author[d]{L. Merten}
\author[a,b]{J. Becker Tjus}
\author[a,b]{H. Fichtner}
\author[e,f]{M. J. Pueschel}
\author[g,h]{E. G. Zweibel}

\affiliation[a]{Theoretical Physics IV: Plasma-Astroparticle Physics, Faculty for Physics and Astronomy, Ruhr-University Bochum, 44780 Bochum, Germany}
\affiliation[b]{Ruhr Astroparticle and Plasma Physics Center (RAPP Center), Ruhr-University Bochum, 44780 Bochum, Germany}
\affiliation[c]{IRFU, CEA Université Paris-Saclay, F-91191 Gif-sur-Yvette, France}
\affiliation[d]{Institute for Astro- \& Particle Physics, University of Innsbruck, 6020 Innsbruck, Austria}
\affiliation[e]{Dutch Institute for Fundamental Energy Research, 5612 AJ Eindhoven, The Netherlands}
\affiliation[f]{Eindhoven University of Technology, 5600 MB Eindhoven, The Netherlands}
\affiliation[g]{Department of Physics, University of Wisconsin-Madison, Madison, WI 53706, U.S.A.}
\affiliation[h]{Department of Astronomy, University of Wisconsin-Madison, Madison, WI 53706, U.S.A.}

\emailAdd{jdo@tp4.rub.de}

\FullConference{%
  *** The European Physical Society Conference on High Energy Physics (EPS-HEP2021), ***\\
  *** 26-30 July 2021 ***\\
  *** Online conference, jointly organized by Universität Hamburg and the research center DESY ***
}
\abstract{
The propagation of cosmic rays can be described as a diffusive motion in most galactic environments. High-energy gamma-rays measured by Fermi have allowed inference of a gradient in the cosmic-ray density and spectral energy behavior in the Milky Way, which is not predicted by models. Here, a turbulence-dependent diffusion model is used to probe different types of cosmic-ray diffusion tensors. Crucially, it is demonstrated that the observed gradients can be explained through turbulence-dependent energy-scaling of the diffusion tensor.

}

\begin{document}
\maketitle
\input{01-Introduction}

\input{02-MilkyWay}
\input{03-Timescale}
\input{04-Simulation}

\input{05-Conclusion}

\input{Bib}
\end{document}

%% file: 01-Introduction.tex
\section{Introduction}
In many astrophysical environments the transport of charged Cosmic Rays (CRs) is described by a diffusion-advection approximation\cite{Strong98, Evoli08, Kissmann14}. 
Here, transport of Galactic CRs is usually described by the Parker transport equation
\begin{align}
    \frac{\partial n}{\partial t} + \vec{u}\cdot\nabla n &= \nabla\cdot(\hat{\kappa}\nabla n) + \frac{1}{p^2}\frac{\partial}{\partial p}\left(p^2\kappa_{pp}\frac{\partial n}{\partial p}\right)  + \frac{p}{3}(\nabla\cdot \vec{u})\frac{\partial n}{\partial p} + S\quad . \label{eq:ParkerTransport}
\end{align}
Here, $\vec{u}$ denotes the advection speed, $p$ the momentum, $n$ the particle distribution, $\hat{\kappa}$ the spatial diffusion tensor, $\kappa_{pp}$ the scalar momentum diffusion coefficient and $S$ the source terms. For this equation isotropic momentum diffusion is assumed and all types of losses due to interaction are neglected.  

Depending on the characteristic timescales of the systems, transport is typically either dominated by spatial diffusion or advection. In most environments, diffusion plays an important role, and it is therefore of high importance to develop a fundamental understanding of the diffusion tensor.
For a field that can be decomposed in a regular component $\vec{B}$ and a turbulent component $\vec{b}$, the diffusion tensor becomes diagonal in a local field-align coordinate system to $\hat{\kappa} = \mathrm{diag}(\kappa_\perp, \kappa_\perp, \kappa_\parallel)$, when antisymmetric diffusion coefficients are negligible \cite{Kopp2012}. The eigenvalues of this tensor can be determined in test particle simulations for different magnetic field configurations as done in \cite{Reichherzer2020, Reichherzer2021}. We are using these results for a description of the diffusion tensor (see Section \ref{sec:2}).

\begin{figure}
\centering 
\includegraphics[width=\textwidth]{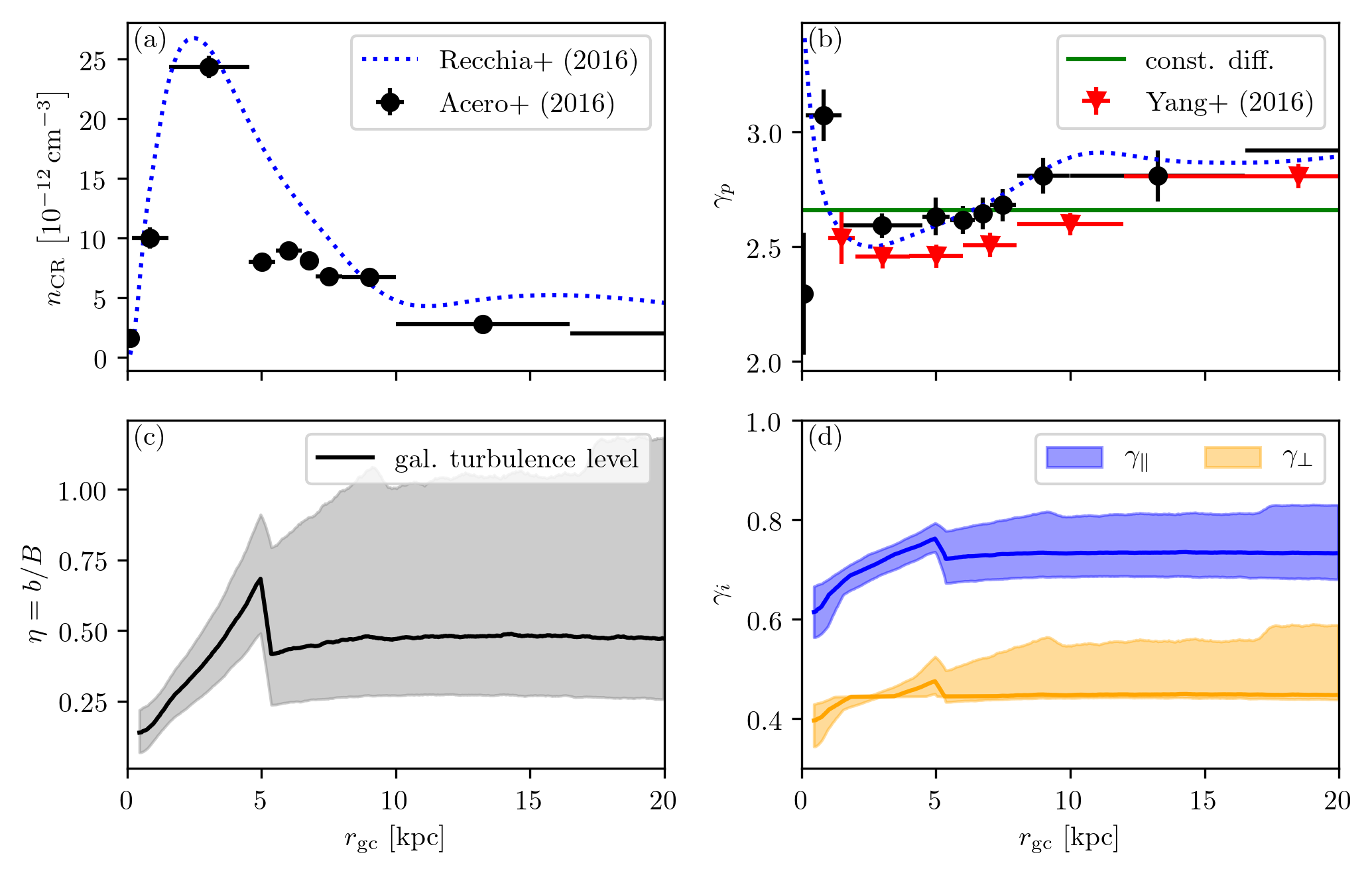}
\caption{\textit{Panel (a) and (b):} Measurements of the cosmic-ray gradient in the Milky Way \cite{Acero2016, Yang2016} in different observables. Panel (a) shows the cosmic ray density profile $n_\mathrm{CR}$ and panel (b) the spectral proton index. Additionally the prediction for a constant diffusion (green solid line) and the fit from Recchia et al.~\cite{Recchia2016} for the cosmic ray density and the spectral index (blue dotted line) is shown. 
\textit{Panel (c):} Turbulence level of the combined magnetic field model of \cite{Kleimann2019} and \cite{Guenduez2020}. The gray band indicates the inter quartil range.  
\textit{Panel (d):} Derived spectral index (this work) for the parallel (blue) and perpendicular (orange) diffusion coefficent.} \label{fig:GalExcess}
\end{figure}

The observations of the radial changes in the cosmic-ray spectra and density in the Milky Way observed by Fermi-LAT \cite{Acero2016, Yang2016}, known as the \textit{Galactic Gradient Problem}, are shown in Fig.~\ref{fig:GalExcess}(a and b). The comparison of the prediction by a spatial constant diffusion (green solid line) and the observed data by Acero et al.~\cite{Acero2016} (black circles) and by Yang et al.~\cite{Yang2016} (red triangles) indicate that the diffusion cannot be spatially constant.

%% file: 02-MilkyWay.tex
\section{Turbulence-dependent diffusion in the Milky Way}\label{sec:2}
\begin{figure}[ht]
\begin{minipage}{0.45\textwidth}
\centering
\includegraphics[width=\textwidth]{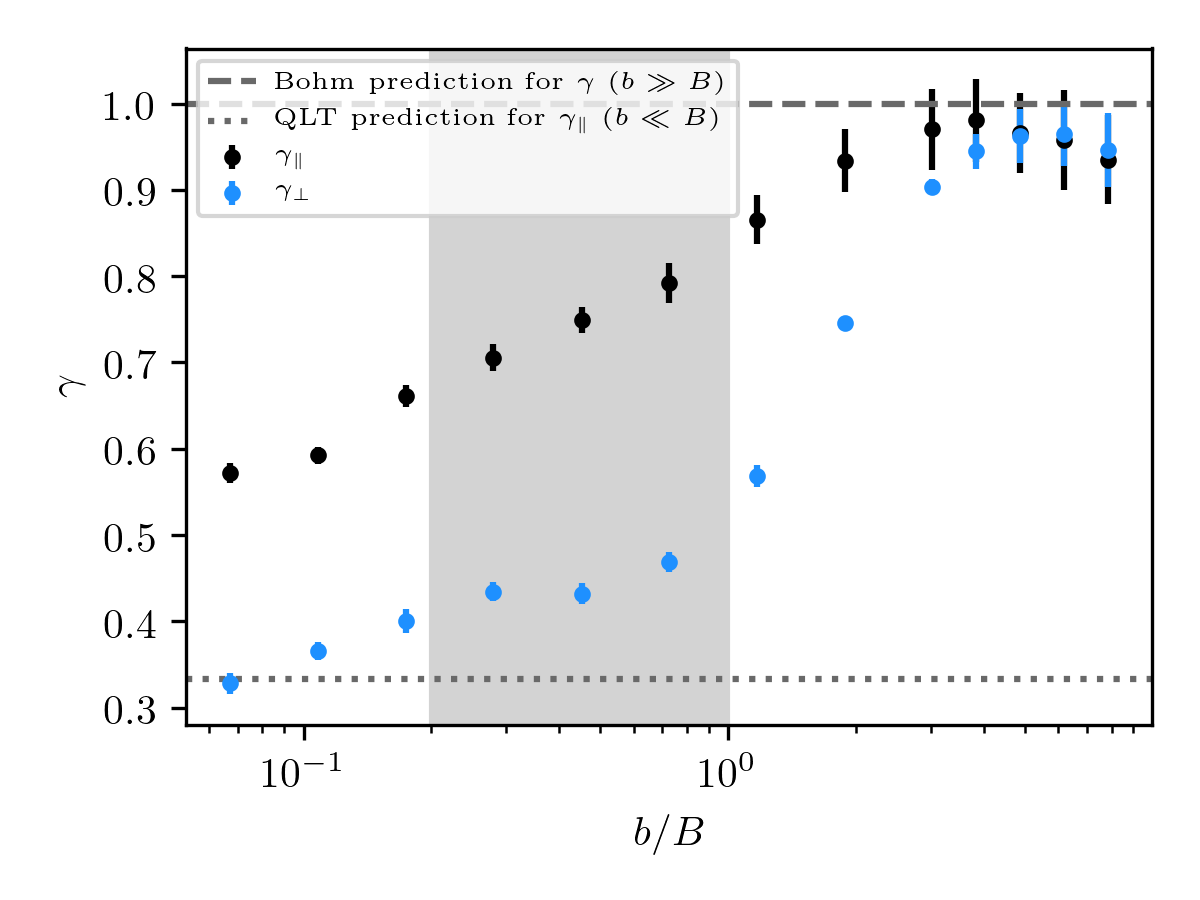}
\caption{Turbulence dependence of the spectral index of the diffusioncoefficent. Data taken from \cite{Reichherzer2021}. The highlighted gray band corresponds to the Galactic turbulence level.} 
\label{fig:gamma}
\end{minipage}
\hfill
\begin{minipage}{0.45\textwidth}
\includegraphics[width=\textwidth]{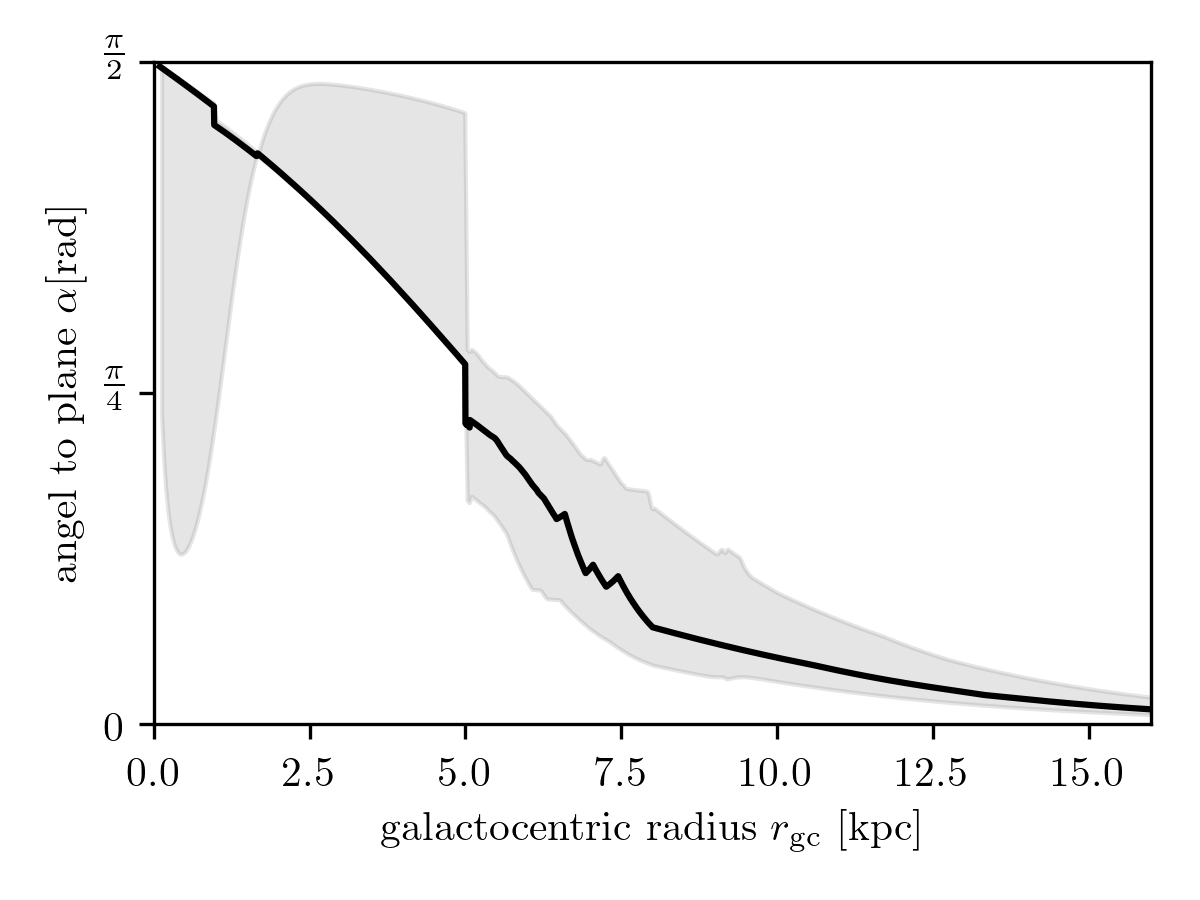}
\caption{Mean angle between the magnetic field lines and the Galactic plane as a function of the galactocentric radius. The magnetic field is a superposition of \cite{Kleimann2019} and \cite{Guenduez2020}.}
\label{fig:Winkel}
\end{minipage}

\end{figure}

In the analysis of the dependencies of the diffusion coefficients done in \cite{Reichherzer2020, Reichherzer2021} for a charged cosmic ray particle in a combination of a large-scale magnetic field $\vec{B}$ and a synthetic, Kolomogerov-like turbulent field $\vec{b}$, it has been shown that the spectral index $\gamma_i$ of the energy dependence of the diffusion coefficient $\kappa_i \propto E^{\gamma_i}$, where $i \in \{ \parallel, \perp\}$, depends on the turbulence level $\eta = b/B$. 
Figure \ref{fig:gamma} shows the results for the parallel and perpendicular component. 
The results are compatible with QLT, as the expected energy behavior of $E^{1/3}$ is approached toward small values of $\eta$ and toward high values, reaching the Bohm limit $E^{1}$ for the parallel component.

Therefore, a detailed knowledge of the turbulence level in the Milky Way is necessary. Here we adopt the global model of \cite{Kleimann2019}, which is a modification of the model by \cite{JF12}. Due to the fact that the coherent component for the Galactic Center is neglected we use a superposition with the inter cloud component of \cite{Guenduez2020}. The resulting turbulence level is displayed in Fig.~\ref{fig:GalExcess} (c). 

Combining the turbulence level as measured in the Milky Way with the turbulence dependent spectral index of the diffusion coefficient leads to a radial gradient in the spectral shape, at least for the inner part of the galaxy (see Fig.~\ref{fig:GalExcess} (d) and \cite{Reichherzer2021}). 

%% file: 03-Timescale.tex
\section{Timescale analysis}
As a first step to simplify our model we neglect momentum diffusion $\kappa_{pp} = 0 $ and adiabatic effects $\nabla \cdot \Vec{u} = 0$ and search for a stationary solution. In this case Eq. \ref{eq:ParkerTransport} simplifies to 
\begin{equation}
    0 \approx 
    \frac{\partial n}{\partial t} = 
    S + \nabla \cdot (\hat{\kappa} \nabla n) - \Vec{u} \cdot \nabla n \quad .
\end{equation}
Assuming that the sources of Galactic CRs are from the same type, radial changes in the spectral shape can only come from the diffusion term. This term can be approximated by using the effective escape distance $d_\parallel$ and $d_\perp$ as
\begin{equation}
    \nabla ( \hat{\kappa} \nabla n) \approx \left( \frac{\kappa_\parallel}{d_\parallel^2} + \frac{\kappa_\perp}{d_\perp^2}\right)n  = - \left( \frac{1}{\tau_\parallel} + \frac{1}{\tau_\perp} \right)n \sim - \frac{n}{\tau_\mathrm{diff}} \, .
\end{equation}
Here $\tau_i = d_i^2/\kappa_i$, with $i\in\{\parallel, \perp\}$ are the individual escape times. The diffusion time scale is dominated by the shortest escape time $\tau_\mathrm{diff} \sim \min(\tau_\parallel, \tau_\perp)$. 

As the Galactic height is much smaller than the radius, the escape direction is given by the orientation of the field lines at given galactocentric radius. In Fig. \ref{fig:Winkel} the mean angle between the magnetic field line and the Galactic plane as a function of the galactocentric radius is given. It can be seen, that in the inner part of the Galaxy ($r_\mathrm{gc} < 5 \, \mathrm{kpc}$) the field lines mainly lie in a direction that is perpendicular to the Galactic plane. As this is also the shortest way out of the system, escape should be along the field lines, so dominated by parallel diffusion. In the outer part of the Galaxy ($r_\mathrm{rg}> 5 \, \mathrm{kpc}$) the field lines lie in the Galactic plane. 
As escape should be happening in the perpendicular direction, it is expected to be dominated by perpendicular diffusion.
Only at the edge of the Galaxy ($r_\mathrm{rg} \gtrsim 19 \, \mathrm{kpc}$), where escape along the galactocentric radius becomes possible, radial escape via parallel diffusion can be dominant.

For the diffusion indices we use constant values, as the variation along the Galactic radius is rather weak as shown in Fig.~\ref{fig:GalExcess}(d). Here, we take $\gamma_\parallel \sim 0.7$ and $\gamma_\perp \sim 0.4$. 
Using such rather strong energy dependencies for parallel escape results in a steep spectrum in the inner Galaxy,
$n \propto E^{-\gamma_s - \gamma_i} \sim E^{-2.9 \pm 0.1}$, using a source index $\gamma_s \sim 2.3 \pm 0.1$. In the outer Galaxy we get $n\propto E^{-2.7 \pm 0.1}$ and at the edge $n\propto E^{-2.9 \pm 0.1}$. Comparing these values with the observation (see Fig.~\ref{fig:GalExcess}(b)) we only fit the second datapoint by Acero et al.~\cite{Acero2016} in the inner part. 
A flat spectrum as observed in the inner Galaxy would result from advective-dominated transport, which has already been discussed in \cite{Evoli08, Everet2008, Everet2010}. A first conclusion from these analytical estimates is therefore that advection is espected to dominate the transport in the inner Galaxy, perpendicular transport starts to dominate at 5kpc and outward, resulting in a steeper spectrum, and finally going over to parallel transport, steepening the spectrum even further.

%% file: 04-Simulation.tex
\section{Simulation}
To estimate the implications of the turbulence dependence of the diffusion coefficient in a full 3D-model we perform several simulations with different configurations of the diffusion tensor. 
\begin{figure}[htb]
\centering
\includegraphics[width=0.95\textwidth]{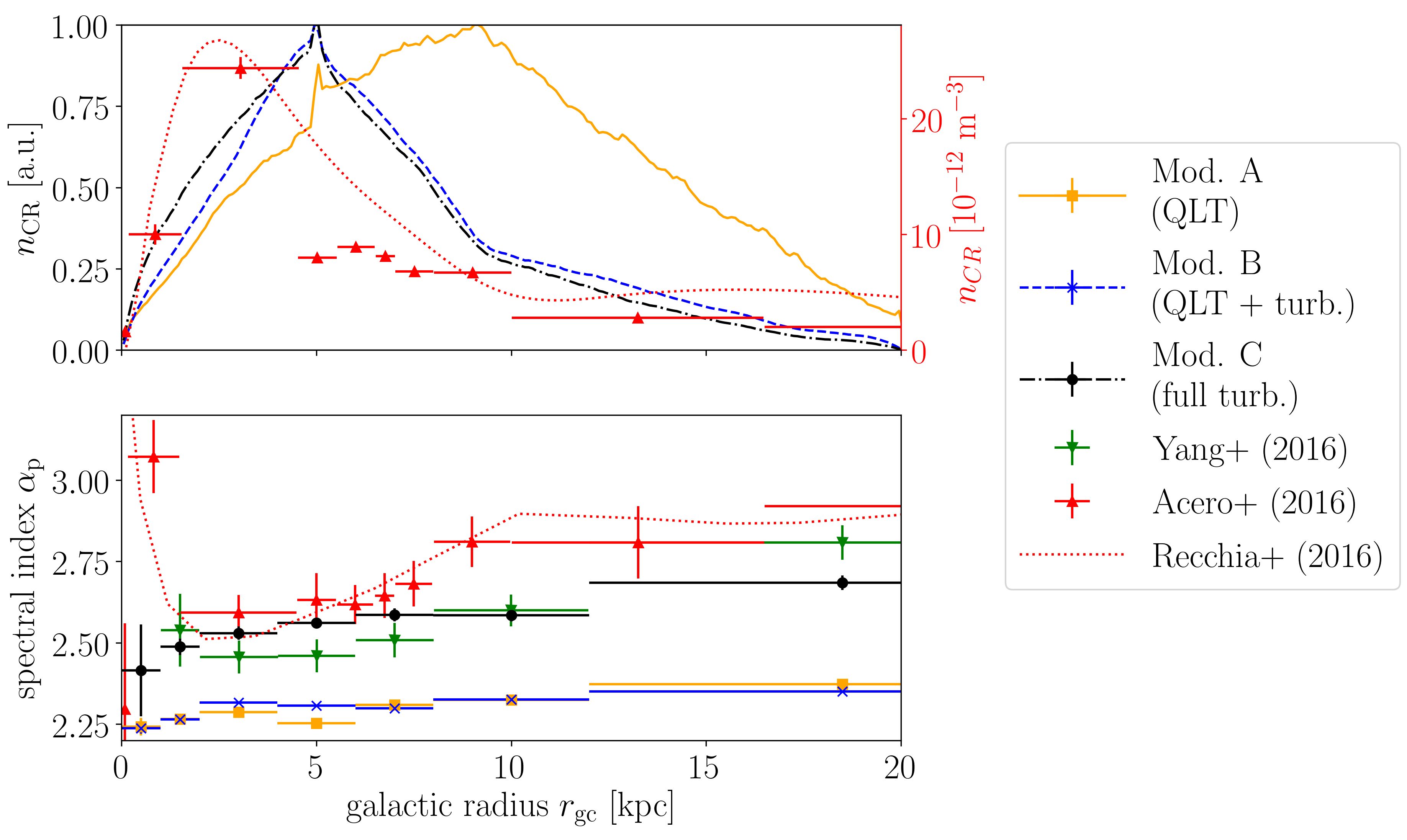}
\caption{Simulation results for turbulent dependent diffusion. The upper panel shows the cosmic-ray density and the lower one the spectral index.}
\label{fig:result}
\end{figure}

\paragraph{General setting}
We use CRPropa, which includes the possibility to propagate diffusively via the method of stochastic differential equation since version 3.1 \cite{Merten2017}. We further optimized the code locally to also work in the limit of particle velocities v<c. 
The simulation is done with protons in the energy range form 50 GeV to 100 TeV, which corresponds to the primary energy producing photons observed by the Fermi-LAT. The source position follows the SNR distribution given in \cite{Merten2017}. We use the combined magnetic field model of \cite{Kleimann2019} and \cite{Guenduez2020} as described before. As a main energy loss for the hadronic particles proton-proton interactions are taken into account using the parameterization of the cross section given in \cite{Kamae2006} and a gas distribution as implemented in the newest version of CRPropa, 3.2 \cite{CRPropa32}. We perform the simulations for $2\cdot 10^5$ pseudo-particles and calculate a stationary solution from the time-dependent result as described in \cite{Merten2017}.

\paragraph{Diffusion coefficients}
We compare three different types of diffusion coefficients. The first model (A) corresponds to the standard approach in \cite{Merten2017} for the quasi-linear-theory (QLT). The diffusion tensor is anisotropic with a constant ratio $\epsilon = \kappa_\perp / \kappa_\parallel = 0.1$. The parallel component follows
\begin{equation}
    \kappa_\parallel^A (E) = \kappa_0 \cdot \EGeV ^\frac{1}{3} \quad, 
\end{equation}
where $\EGeV$ denotes the particle energy in units of 4 GeV and $\kappa_0 = 6.1 \cdot 10^{24} \,  \mathrm{m}^2\, \mathrm{s}^{-1}$ is the observed value at Earth \cite{BeckerTjus2020}. In a second model (B) we take the predicted turbulence scaling into account. The diffusion coefficent is given as 
\begin{equation}
    \kappa_\parallel^B (E, \eta) = \kappa_0 \cdot \eta^{-2} \eta_\odot^2 \cdot \EGeV^\frac{1}{3} \quad \mathrm{and} \quad  \kappa_\perp^B(E, \eta) = \kappa_0 \left( \eta \, \eta_\odot \right)^2 \cdot \EGeV^\frac{1}{3}
\end{equation}
with $\eta$ as the turbulence level and $\eta_\odot$ as the turbulence at Earth. In this case the ratio between perpendicular and parallel components become $\epsilon = \eta^4$ which is predicted by the QLT.
In the third model (C) the full turbulence dependence is taken into account. Here, the spectral index $\gamma_i = \gamma_i(\eta)$ is interpolated for the spatial dependent turbulence level. The diffusion coefficient follows as
\begin{equation}
    \kappa_\parallel^C (E, \eta) = \kappa_0 \cdot 
    (\eta / \eta_\odot)^{-2}
    \cdot \EGeV ^{\gamma_\parallel(\eta)} \quad \mathrm{and} \quad  \kappa_\perp^C (E, \eta) = \kappa_0 \left( \eta\, \eta_\odot \right)^2 \cdot \EGeV^{\gamma_\perp (\eta)} \, . 
\end{equation}

\paragraph{Results}
The results of the simulation are shown in Fig.~\ref{fig:result}. We binned the radial position of the simulated pseudo-particles and took their number density, which is proportional to the CR density. The upper panel provides a qualitative comparison between the shape of the observed and simulated density. Considering only the energy scaling of the diffusion coefficient (model A) leads to an excess of the CR density at greater radii compared to the turbulence-scaled diffusion coefficients (model B or C). However, the turbulence dependent diffusion scenarios peaks at larger radii compared to the data. 
Comparing the spectral behavior (see Fig.~\ref{fig:result}(b)) it only shows minor discrepancy between models A and B, due to the fact that both cases have the same energy scaling of the diffusion coefficient. Both lines show only marginal deviation from a radial constant index. This deviation may result from geometrical effects of the averaged field line direction at given galactocentric radius. The model C shows a better agreement with the data although even here the gradient is much smaller as compared to data.

%% file: 05-Conclusion.tex
\section{Conclusion}
The turbulence dependence of the diffusion tensor plays an important role for the explanation of the observation of the so-called galactic excess.
A detailed modeling of the turbulence has direct influence on the prediction of the cosmic-ray density and the spectral index of the cosmic-ray spectrum.